\documentclass[runningheads]{llncs}
\usepackage[T1]{fontenc}

\usepackage{graphicx}
\usepackage{hyperref}
\usepackage{color}

\usepackage{colortbl}
\usepackage[most]{tcolorbox}
\usepackage{booktabs}
\usepackage{caption}
\usepackage{wrapfig}
\usepackage{enumitem}

\newcommand{\code}[1]{{\small\ttfamily#1}}
\newcommand{\zb}{{\small\textsf{zbMATH Open}}~}

\usepackage{listings}
\usepackage{xcolor}
\definecolor{eclipseStrings}{RGB}{42,0.0,255}
\definecolor{eclipseKeywords}{RGB}{127,0,85}
\definecolor{backcolour}{RGB}{245,248,250}
\definecolor{codegray}{rgb}{0.5,0.5,0.5}
\colorlet{numb}{magenta!60!black}
\lstdefinelanguage{json}{
    basicstyle=\footnotesize\ttfamily,
    commentstyle=\color{eclipseStrings}, %
    stringstyle=\color{eclipseKeywords}, %
    numbers=left,
    numberstyle=\tiny\color{codegray},
    numbersep=2pt,                  
    stepnumber=1,
    showstringspaces=false,
    breaklines=true,
    frame=lines,
    backgroundcolor=\color{backcolour},
    string=[s]{"}{"},
    comment=[l]{:\ "},
}

\definecolor{tagcolor}{RGB}{127,0,85}        
\definecolor{valuecolor}{RGB}{0,0,180}      
\definecolor{backcolour}{RGB}{245,248,250}  
\definecolor{commentgray}{rgb}{0.5,0.5,0.5} 

\lstdefinelanguage{xml}{
    basicstyle=\footnotesize\ttfamily,
    showstringspaces=false,
    breaklines=true,
    frame=lines,
    backgroundcolor=\color{backcolour},  
    morestring=[b]",
    stringstyle=\color{valuecolor},                  %
    moredelim=[s][\color{tagcolor}]{<}{>},           %
    moredelim=[s][\color{tagcolor}]{</}{>},          %
    comment=[s]{<!--}{-->},                      
    commentstyle=\color{codegray}\itshape,
}

\newcommand{\orcid}[1]{\href{https://orcid.org/#1}{\includegraphics[width=10pt]{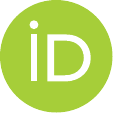}}}

\begin{document}

\title{The \textsf{zbMATH Open} Knowledge Graph: Tracing Centuries of Mathematical Research}

\titlerunning{The \textsf{zbMATH Open} Knowledge Graph}

\author{Yuni Susanti{\orcid{0009-0001-1314-0286}} \and
Moritz Schubotz{\orcid{0000-0001-7141-4997}}}
\authorrunning{Y. Susanti et al.}

\institute{FIZ Karlsruhe, Berlin, Germany\\
\email{\{yuni.susanti, moritz.schubotz\}@fiz-karlsruhe.de}}

\maketitle              %
\begin{abstract}
\label{firstpage}
We present \zb Knowledge Graph, a large-scale RDF knowledge graph (KG) covering more than 250 years of mathematical scholarship. Unlike existing scholarly knowledge graphs that primarily capture bibliographic metadata and citation structures, the \zb KG integrates expert-curated semantic content, including reviews, keywords, subject classifications, software references, and disambiguated authorship. 
This combination of domain-specific representation of mathematical knowledge and extensive temporal coverage supports analyses that require fine-grained exploration of mathematical concepts, research fields, and scholarly relationships over time.
The resulting graph comprises 34 million entities and 168 million RDF triples represented using established Semantic Web vocabularies, supporting interoperability and FAIR data principles. 
We further demonstrate its capabilities through query-driven \textit{historically grounded scholarly exploration} use cases, illustrating how the knowledge graph can surface relationships and patterns that may be difficult to identify from bibliographic and citation information alone.
The \zb KG provides an open semantic infrastructure for studying the development of mathematical knowledge and tracing scholarly connections across centuries of scholarship.
\vspace{3mm}
\\
[0.2cm]
{\scriptsize
\begin{tabular}{ll}
    \textbf{Resource type: }& Dataset/Knowledge Graph \\ 
    \textbf{License: }&\href{https://creativecommons.org/licenses/by-sa/4.0/}{CC-BY-SA 4.0}\\ 
    \textbf{DOI: } & \url{https://doi.org/10.5281/zenodo.21497975}\\
     \textbf{URL:}&\url{https://github.com/zbMATHOpen/zbmath-open-kg} \\
\end{tabular}
}

\keywords{knowledge graphs \and scholarly data \and mathematics}
\end{abstract}

\vspace{-4mm}
\section{Introduction}
\label{sec:intro}
\vspace{-2mm}
Mathematics is one of the oldest continuously evolving scientific disciplines, with a documented scholarly record spanning over many centuries. Beyond formal results, its literature captures the long-term conceptual, methodological, and intellectual development of mathematical knowledge. Despite the availability of millions of digitized publications, the semantic structure and historical development of mathematical knowledge remain difficult to analyze computationally. Existing platforms primarily support keyword-based search and citation-based discovery, offering limited support for investigating how mathematical concepts emerge, how ideas migrate across subfields, and how intellectual lineages are formed across generations.
\cite{Chebukov_2013,global_math_library_2014,elizarov2017ontomathdigitalecosystemontologies}.

In this work, we present \zb Knowledge Graph (\zb KG), a large-scale and historically comprehensive RDF-based knowledge graph of mathematical publications and scholarly knowledge. Constructed from the \zb platform~\cite{zbmathFacts}, the longest-running abstracting and reviewing service in mathematics, the resource builds upon more than 150 years of curated mathematical scholarship, covering over four million publications spanning 250 years~\cite{zbmathFacts}. Unlike existing scholarly knowledge graphs that primarily represent bibliographic metadata and citation networks~\cite{openalex,semopenalex,arxiv,dblp,semantic_scholar}, \zb KG integrates expert-curated semantic knowledge, including:

\begin{itemize}
\item expert-written reviews providing contextual interpretations of publications,
\item disambiguated identities of scholars, including authors and reviewers,
\item expert-curated controlled mathematical keywords,
\item expert-assigned Mathematics Subject Classification (MSC)~\cite{msc2020} codes, a domain-specific taxonomy that has provided a stable conceptual organization of mathematics across decades,
\item software references associated with mathematical publications.
\end{itemize}

To support interoperability within Semantic Web ecosystem, we construct the knowledge graph using established vocabularies and ontologies.
The resulting graph comprises more than 34 million entities and 168 million RDF triples and is designed to align with FAIR~\cite{fair} principles. 
Beyond data publication, the \zb KG supports distinct use cases of scholarly knowledge discovery we collectively term \textit{historically grounded scholarly exploration}: the analysis of long-term intellectual relationships and scholarly development beyond citation networks and 
bibliographic metadata. By combining expert-curated semantics with centuries of scholarly records, the \zb KG establishes a foundation for studying the development of mathematical knowledge and tracing intellectual relationships across centuries of scholarly development. \\

\noindent \textbf{Contribution.} Our main contributions can be summarized as follows:
\begin{enumerate}[itemsep=4pt]
    \item We construct \zb KG, a large-scale RDF knowledge graph spanning over 250 years of mathematical scholarship grounded in expert-curated semantic contents. The resource 
    is publicly released and made accessible\footnote{See Page~\pageref{firstpage} for the DOI and URL; SPARQL endpoint is available on GitHub.} through the following:
        \begin{enumerate}
        \item We provide the full RDF data dumps via Zenodo for long-term accessibility and preservation. The dump is updated periodically, approximately twice per year, subject to the availability of upstream data sources.
        \item We deploy the knowledge graph in a triple store and provide a public SPARQL endpoint for direct querying.
        \item We enable URI resolution for the core entities (e.g., publications, scholars, software references), supporting linked data access.
        \item We release the full source code for the knowledge graph construction to support reproducibility, customization and future extensions.
        \end{enumerate}
        
    \item We design the resource according to FAIR principles and Semantic Web standards.
    The OWL ontology is published with VoID, DCAT and PROV-O metadata description to support machine-readable discovery and reuse.
    
    \item We evaluate the resource at three complementary levels: serialization validity, structural consistency, and competency-question coverage. Beyond these validations, we demonstrate the practical utility of the resource through query-driven historically grounded scholarly exploration use cases.
\end{enumerate}

\vspace{-2mm}
\section{Related Work}
\label{sec:relwork}
\vspace{-2mm}
Knowledge graphs have become an important paradigm for representing and interlinking scholarly information. Large-scale initiatives such as the Microsoft Academic Graph (MAG)~\cite{microsoftacademic}, its Linked Data transformation Microsoft Academic Knowledge Graph (MAKG)~\cite{makg}, and its successor OpenAlex~\cite{openalex} provide extensive coverage of scholarly metadata and citation networks. Semantic representations such as SemOpenAlex~\cite{semopenalex} further improve interoperability with Semantic Web technologies, while domain-specific resources such as LPWC~\cite{lpwc23} focuses on machine learning publications, and SemRepo~\cite{semrepo} extend the LPWC scholarly knowledge graphs by linking publications with software and research artifacts. Similarly, established scholarly databases including arXiv~\cite{arxiv}, DBLP~\cite{dblp}, and Semantic Scholar~\cite{semantic_scholar} provide broad bibliographic coverage.
However, these resources primarily represent publication-level metadata and citation relationships, with limited 
expert-curated, high-quality semantic knowledge and long-term conceptual development within a scientific domain.

These projects cover a wide spectrum of scholarly disciplines; however their representation and coverage of mathematical publication and knowledge remains limited. Several initiatives have been undertaken to capture the structure and semantics of mathematical knowledge. For instance, the OntoMath ontology~\cite{ontomath} models mathematical concepts and relations for semantic search, while projects such as Formal Abstracts~\cite{formalabstracts}, Lean~\cite{leanmathlib}, and Coq~\cite{coqmathcomp} focus on formalizing mathematical statements and machine-verifiable knowledge. MMLKG provides a thesaurus for describing mathematical definitions, statements, and proofs based on the Mizar Mathematical Library~\cite{mmlkg}. These efforts provide deep representations of mathematical content but primarily target formal reasoning. 
More recently, AutoMathKG~\cite{automath} integrates paper with theorems and proofs; however, its construction relies heavily on semi-automatic extraction and LLM-assisted processing and focused to enhance LLM mathematical reasoning. MaRDI knowledge graph~\cite{schubotz2023bravomardiwikibasepowered}, based on Wikidata data modeling, integrates mathematical research data from multiple sources, 
offering broad coverage of mathematical research artifacts. 
However, their structure is tightly coupled to Wikidata’s internal model, relying on Wikibase identifiers and property models rather than standard Semantic Web vocabularies~\cite{schubotz2023bravomardiwikibasepowered}. 

The \zb KG complements these efforts by providing an RDF-native representation of mathematically curated scholarly knowledge. Unlike Wikidata-based approaches such as MaRDI, 
the \zb KG is built on established Semantic Web vocabularies and ontologies and designed to align to its standards and FAIR principles.
Additionally, the explicit modeling of expert-curated semantic contents enables analyses requiring fine-grained understanding of mathematical concepts, fields, and intellectual relationships.

\vspace{-2mm}
\section{The \textsf{zbMATH Open Knowledge Graph}}
\label{sec:kg}
In this section, we describe the construction process of the \zb KG (\S\ref{sec:engineering}), followed by ontology and schema design (\S\ref{sec:ontology}) and key statistics (\S\ref{sec:statistics}). Finally, we evaluate the graph to assess its validity and coverage (\S\ref{sec:evaluation}).

\vspace{-2mm}
\subsection{Knowledge Graph Construction}
\label{sec:engineering}
The development of the \zb KG follows a knowledge engineering methodology largely guided by METHONTOLOGY~\cite{methonto97}, complemented by ontology engineering practices from SAMOD~\cite{samod} and LOT~\cite{lotmet}. 
The construction process consists of the following stages:

\begin{enumerate}[itemsep=0.5em]
    \item \textbf{Requirements Specification.} We first defined the scope of the knowledge graph by identifying the core entities, relationships, and intended usage scenarios. Competency Questions (CQs) were formulated and checked by human experts to capture representative scholarly information needs, including historical semantic exploration. 
    We share the full CQs in our repository.

    \item \textbf{Knowledge Acquisition.} The raw dataset was acquired from the \zb platform via its official API\footnote{\url{https://oai.zbmath.org/} December 2025 snapshot (Accessed: 15/12/2025)}. The acquired metadata served as the primary source for ontology development and instance generation.

    \item \textbf{Conceptualization.} Based on the acquired data, we iteratively developed the conceptual model and structural requirement of the knowledge graph. Core classes such as publications, authors, and reviews
    were identified together with their semantic relationships, including the required properties and datatypes. The conceptual model was repeatedly refined as additional metadata characteristics and competency questions were analyzed.

    \item \textbf{Vocabulary Integration.} We maximized interoperability through the reuse of established Semantic Web vocabularies. For instance, scholarly entities are primarily modeled using \textit{schema.org}~\cite{guha2016schema},
    citation relations using \textit{cito}~\cite{peroni2012cito}, and concept organization using \textit{skos}~\cite{miles2009skos}. Domain-specific concepts 
    (e.g., mathematical subject classification and keywords) are introduced within the zbMATH namespace. See the detail of ontology and schema design in~\ref{sec:ontology}.

    \item \textbf{Implementation.} The pipeline automatically converts the raw data into RDF triples.
    To facilitate customization (e.g., custom KG subsets generation), we share the full source code in our repository. 

    \item \textbf{Evaluation and Maintenance.} The generated RDF knowledge graph is evaluated at three levels to assess its validity and coverage: serialization validity, structural consistency, and competency-question coverage.
    See the detailed evaluation in~\ref{sec:evaluation}. Maintenance is planned through periodic updates with versioned release, supported by the pipeline that enables regeneration from the \zb snapshots as new releases become available.
\end{enumerate}

\subsection{Ontology and Schema Design}
\label{sec:ontology}
Figure~\ref{fig:onto} illustrates the schema design of the \zb KG, including an overview of the entity types, the object properties, and the data type properties. To support interoperability with the Linked Data ecosystem, the model follows the best practice of semantic reuse strategy by adopting established Semantic Web vocabularies wherever possible and extending them with \zb-specific classes and persistent identifiers for mathematical resources. 

{
  \setlength{\textfloatsep}{6pt} %
  \setlength{\floatsep}{5pt}      %
  \setlength{\intextsep}{6pt}     %
\begin{figure}
    \centering
    \includegraphics[width=1\linewidth]{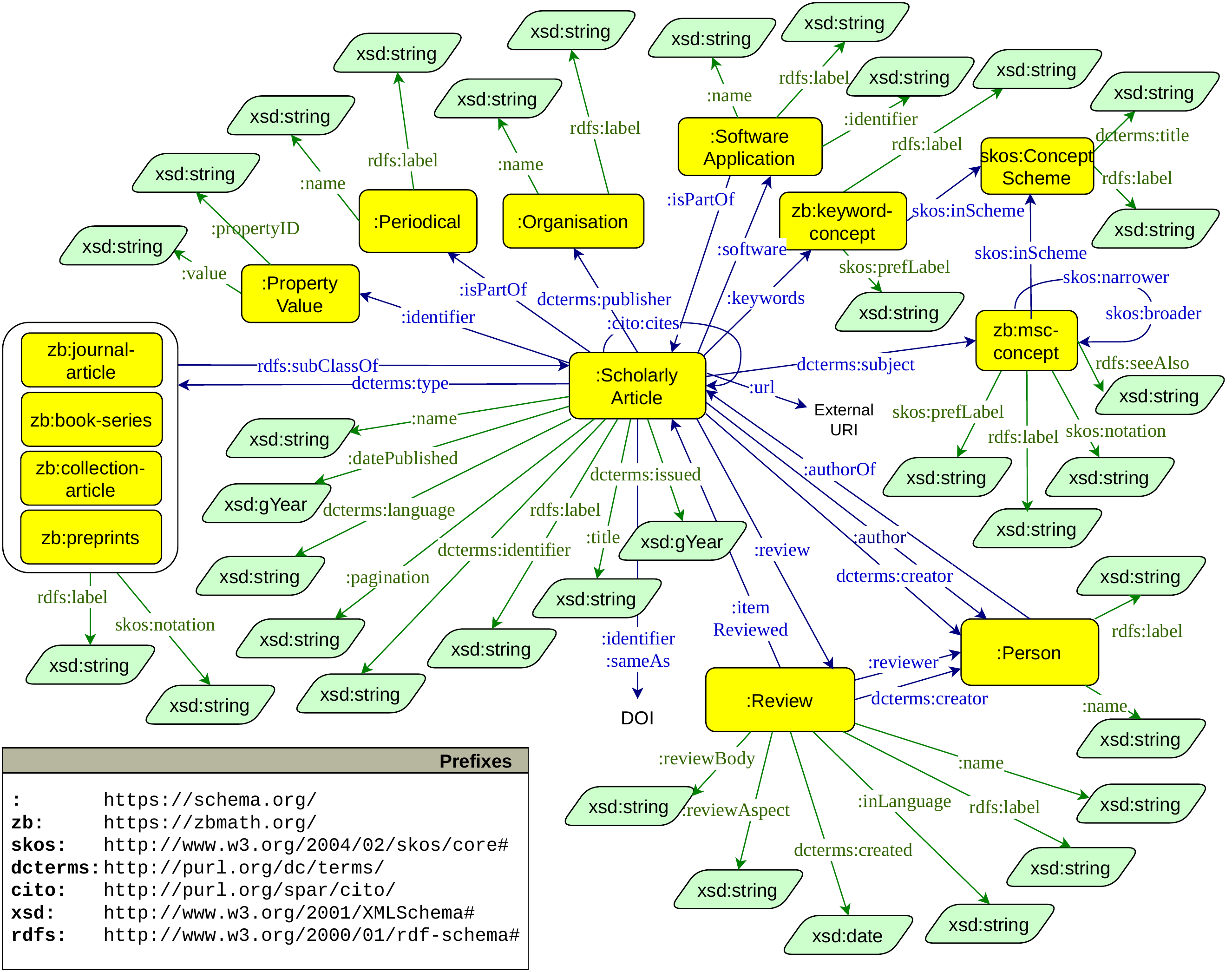}
    \caption{Schema overview of the \zb Knowledge Graph.\protect\footnotemark}
    % \caption{Schema overview of the \zb Knowledge Graph.\footnotemark}
    \label{fig:onto}
\end{figure}
}
\footnotetext{For readability, only the core classes and relationships are shown; the figure is \underline{\textbf{not}} intended as 
an exhaustive representation of all object and properties.}

\paragraph{Scholarly Entity Representation.} The core of the model represents mathematical scholarly publications and their contextual relationships. Publication entities are modeled using \emph{schema.org} terms (\code{schema:})~\cite{guha2016schema}. This representation captures not only the main bibliographic records (\code{schema:ScholarlyArticle}, \code{schema:Person} for representing scholars i.e., authors and reviewers) but also the broader scholarly ecosystem surrounding a scholarly article, including reviews, periodicals, publishers, and associated mathematical software.

\paragraph{Bibliographic and Citation Modeling.} Bibliographic metadata including article titles, publication dates, languages, identifiers (e.g., DOI), url, publishers, and document types are additionally represented using \textit{Dublin Core Terms} (\code{dcterms:})~\cite{dublincore}, following established practices in digital library data modeling. Citation relationships are represented using the \textit{Citation Typing Ontology} (\code{cito:})~\cite{peroni2012cito} for explicit modeling of publication-level citation networks.

\paragraph{Mathematical Semantic Representation.} One of the key features of the \zb KG is the explicit representation of mathematical semantics. The organization of the \textit{Mathematics Subject Classification} (MSC) codes~\cite{msc2020} and expert-curated controlled keywords are modelled using SKOS (\code{skos:})~\cite{miles2009skos}, while retaining persistent \zb-specific identifiers. The hierarchical structure of the MSC taxonomy is modeled using \code{skos:broader} and \code{skos:narrower}, preserving the long-established disciplinary organization of mathematics.

\paragraph{Interoperability and Extensibility.} Additionally, RDF Schema (\code{rdfs:}) is used to define classes, hierarchies, and labels; XML Schema datatypes (\code{xsd:}) provide standardized datatypes for literal values. The ontology defines zbMATH Open-specific classes and persistent IRIs for mathematical resources while maintaining interoperability with the Linked Data ecosystem via established Semantic Web vocabularies. The model is intentionally lightweight and extensible, allowing future integration of additional mathematical resources, such as datasets, formalized results, or mathematical objects (e.g., formulae). The complete ontology specification is publicly available
alongside the released data and machine-readable metadata descriptions.

\subsection{Key Statistics}
\label{sec:statistics}

Table~\ref{tab:kg_metrics} illustrates the scale, coverage, and structural properties of the constructed \zb KG.
With over 34.5 million entities and nearly 170 million relationships spanning 39 distinct predicate types, it represents one of the most comprehensive structured resources in mathematics. On average, each entity is connected to nearly ten others, reflecting a well-linked but non-redundant structure. 
The graph density of $1.45 \times 10^{-7}$ reflects the expected sparsity typical of heterogeneous scholarly RDF graphs, where the number of observed relationships remains small compared to all possible pairwise connections.
{
  \setlength{\textfloatsep}{4pt} %
  \setlength{\floatsep}{4pt}      %
  \setlength{\intextsep}{4pt}     %
\begin{table}[t]
    \centering
    \caption{Quantitative metrics of the \zb Knowledge Graph}
    \label{tab:kg_metrics}
    \begin{tabular}{@{}lll@{}}
    \toprule
    \textbf{Metric} & \textbf{Description} & \textbf{Value}  \\
    \midrule
    Number of Entities & \textit{All distinct nodes (subjects/objects in triples)} & 34,099,406 \\
    Number of RDF Resources & \textit{All distinct nodes (subjects/objects in triples)} & 18,995,799 \\
    Number of Typed Resources & \textit{Resources with \code{:type} declaration} & 15,392,393 \\
    Type Coverage (\%) & \textit{\% of resources with a defined type} & 81.03\% \\
    Number of Edges & \textit{Total number of triples} & 168,670,821 \\ 
    Number of Relation Types & \textit{Unique predicate types used in edges} & 39 \\
    Average Degree & \textit{Mean number of edges per entity} & 9.89 \\
    Graph Density & \textit{Ratio of actual to possible edges} & $1.45 \times 10^{-7}$ \\
    Temporal Coverage & \textit{Time span covered by temporal data} & 1763–2026 \\
    \bottomrule
    \vspace{-3mm}
    \end{tabular}
\end{table}
}

Importantly, 15.4 million RDF entity resources (over 80\%) are explicitly typed, providing semantic annotations that improve interpretability and support more precise queries over bibliographic records.
Although \textsf{zbMATH}, the original platform, was established in 1931~\cite{zbmathFacts}, the knowledge graph captures a temporal span of more than 250 years through the inclusion of publications dating back to 1763. This extensive temporal coverage supports analyses of scholarly production and the development of entities and relationships over time.

Table~\ref{tab:entity_metrics} summarizes the distribution of core entities within \zb KG. The entity distribution reflects the scholarly focus and expert-curated nature of the KG.
At its center are more than 4 million articles, which form the backbone of the graph. These publications are connected to a curated network of over 1.1 million \textit{disambiguated} person entities~\cite{DebZB2024}, including more than 1.12 million authors and nearly 14 thousand reviewers, providing a curated representation of scholarly contributors and supporting reliable attribution of research contributions.
\begin{wraptable}[18]{r}{5cm}\vspace*{-0.25\baselineskip}%
    \centering
    \caption{Core entities in \zb Knowledge Graph}
    \vspace{-7pt}
    \label{tab:entity_metrics}
    \begin{tabular}{@{}lr@{}}
    \toprule
    \textbf{Entity} & \textbf{\# Instances} \\
    \midrule
    Scholarly articles & 4,070,393 \\
    Person & 1,125,474 \\
    ---Authors & 1,125,238 \\
    ---Reviewers & 13,942 \\
    Subjects (MSC) & 6,734 \\
    Keywords & 3,008,881 \\
    Organization & 3,524 \\
    Periodical & 7,350 \\
    Reviews & 3,098,767 \\
    Software & 30,857 \\
    Identifiers & 10,522,258 \\
    ---DOI & 2,446,976 \\
    \bottomrule
    \end{tabular}
\end{wraptable}

The scholarly articles are complemented by 3.10 million expert-written reviews, which provide contextual interpretations and assessments of mathematical publications beyond bibliographic metadata. Furthermore, the presence of 3.01 million controlled keywords and 6,734 mathematics subject classification entities demonstrates the integration of domain-specific semantic annotations that support fine-grained organization and retrieval of mathematical knowledge. The ecosystem is further represented through thousands of organizations i.e., publishers (3,524) and periodicals (e.g., journals) (7,350), as well as over 30,000 references to mathematical software linked to publication, capturing important contextual elements of mathematical research. The resource contains more than 10.5 million identifiers, including nearly 2.5 million DOI references, which facilitate interoperability with external scholarly resources and persistent identification of research outputs. 

\subsection{Evaluation and Validation}
\label{sec:evaluation}
The generated knowledge graph was validated at three complementary levels\footnote{All CQs, SPARQL queries, and codes for evaluation are available in GitHub (Page~\pageref{firstpage}).}: (i) \textit{syntactic validity} of the RDF serialization, (ii) \textit{structural consistency}, and (iii) 
\textit{CQ-based evaluation}, described in detail in the following.

First, the syntactic validity of the generated RDF serialization was verified using both \textit{Apache Jena} and \textit{RDFLib} Pyton library~\cite{rdflib}, and 
was subsequently loaded into the \textit{OpenLink Virtuoso} and \textit{Apache Jena Fuseki} triple stores~\cite{virtuoso} to verify successful ingestion and to support the SPARQL-based validation. All RDF files were successfully parsed and loaded without syntax or ingestion errors, confirming that the generated resource was syntactically valid.

Second, 
the consistency check was performed to assess compliance with the structural requirements defined during the knowledge graph construction. 
These checks targeted various aspects of structural consistency, such as missing required properties, invalid or unexpected datatypes, and inconsistent class assignments, among others. For each requirement, a SPARQL query was executed over the complete graph to retrieve resources violating the corresponding constraint; the number of returned resources was used as the violation count. Representative results from the 
checks performed are summarized in Table~\ref{tab:structural_validation}.

\begin{table}[t]
\centering
\caption{Summary of the structural consistency check results.}
\label{tab:structural_validation}
\begin{tabular}{p{0.25\linewidth} p{0.59\linewidth} r}
\toprule
\textbf{Check} & \textbf{Validation criterion} & \textbf{Violations} \\
\midrule
Required properties & Scholarly articles contain required properties: & 61,380 \\
& \quad \textit{Title} (\texttt{dcterms:title}) & 3 \\
& \quad \textit{Author} (\texttt{dcterms:creator}, \texttt{schema:author}) & 61,377  \\
& \quad \textit{Publication date} (\texttt{schema:datePublished} ) & 0 \\
& \quad \textit{Document type} (\texttt{dcterms:type}) & 0 \\
Datatype consistency
& Date values use the expected RDF datatypes & 0 \\

Author references
& Article authors are represented as \texttt{schema:Person} & 0 \\

Document type
& Article types belong to the defined document-type vocabulary & 0 \\

MSC references
& Article subjects resolve to defined MSC concepts & 0 \\

Keyword references
& Article keywords resolve to defined keyword concepts & 0 \\

Review references
& Reviewed articles and reviewers resolve to resources of the expected classes & 0 \\

SKOS hierarchy
& \texttt{skos:broader} and \texttt{skos:narrower} relations are mutually consistent & 0 \\
\bottomrule
\end{tabular}
\end{table}

The structural check identified 61,380 missing-property violations (3 missing titles, 61,377 missing authors). Manual inspection confirmed that these violations originate from the incomplete records in the source API, where the title or author information is not provided (``\textit{None}''). The issue is reported to zbMATH platform and is expected to be addressed through a future correction of the API. 
All other structural consistency checks resulted in zero violations.

Third, the validity and coverage of the knowledge graph were assessed using CQs, which were defined during development to formalize the information requirements the resource is intended to support. 
A total of 45 CQs were specified, covering the principal retrieval and relational capabilities of the KG, including bibliographic retrieval, scholarly relationships, and temporal and historical information.
We organized the CQs into (i) \textit{Core Bibliographic and Retrieval} and (ii) \textit{Mathematical Scholarly History and Relationships}. 
For each CQ, a SPARQL query was executed against the graph, and the result was manually compared with the expected answer. 
Table~\ref{tab:competency-evaluation} summarizes the evaluation results.

\begin{table}[t]
\centering
\caption{Summary of the CQ-based evaluation results.}
\label{tab:competency-evaluation}
\begin{tabular}{p{0.42\linewidth} c c c c c}
\toprule
\textbf{CQ category} & \textbf{\#CQs} &
\textbf{\textit{Full}} & \textbf{\textit{Partial}} & \textbf{\textit{Not}} & \textbf{Coverage} \\
\midrule
\textit{Core Bibliographic and Retrieval}
& 24 & 23 (95.8\%) & 0 (0\%) & 1 (4.2\%) & \textbf{95.8\%} \\
\textit{Scholarly History and Relationships}
& 21 & 16 (76.2\%) & 4 (19.0\%) & 1 (4.8\%) & \textbf{95.2\%} \\
\bottomrule
\end{tabular}
\end{table}

The CQ-based evaluation demonstrates broad competency coverage, with 95.6\% of evaluated cases being at least \textit{partially-supported} and 86.7\% \textit{fully-supported}. 
The \textit{partially-supported} CQs primarily concern information available through external resources, e.g., Wikidata links that can potentially be used to obtain institutional or country information of the scholar identities. Although such links are available on the platform, they are not currently exposed via the API and thus cannot be incorporated into \zb KG. This limitation can be addressed by exposing these links, which would enable analyses of scholarly relationships across institutions and other biographical information.

The \textit{not-supported} CQs concern information not available in the source, e.g., programming language of mathematical software. While \zb KG links software to publications, it does not currently provide sufficient software-specific metadata to analyze trends in mathematical research infrastructure. Such information would require explicit linkage to external software knowledge graphs (e.g., SemRepo~\cite{semrepo}). Overall,
the evaluation provides an assessment of the current coverage of the resource and to identify areas for future extension.

\vspace{-2mm}
\section{Applications and Use Cases}
\label{sec:use-case-hist}
To demonstrate the utility of the knowledge graph, we present several use case examples we collectively term \textit{historically grounded scholarly exploration}. These use cases illustrate how the distinctive content of the knowledge graph, particularly its extensive temporal coverage, subject classifications based on the MSC taxonomy, expert-curated keywords, and reviewer-author relationships, supports the exploration of mathematical knowledge across generations of scholarship. Importantly, these use cases are \underline{\textbf{not}} intended to establish historical influence; rather, they demonstrate how the KG can surface relationships and patterns for further scholarly and historical investigation. Each use case is operationalized through SPARQL queries, available in our repository.

\subsection{\textit{Precursor} Discovery: Potential Connections Beyond Citation} 
\label{sec:overlooked-precursor}
One potential capability of the \zb KG is to surface historically earlier contributions that may be foundational and semantically relevant to later developments but are not connected through explicit citation. Identifying such potential ``\textit{precursor}'' connections can help reconstruct a more faithful scholarly lineage and provide a richer view of mathematical knowledge development. 

{
\setlength{\textfloatsep}{4pt} 
\setlength{\floatsep}{3pt}      
\setlength{\intextsep}{4pt}     
\begin{figure}
    \centering
    \includegraphics[width=0.85\linewidth]{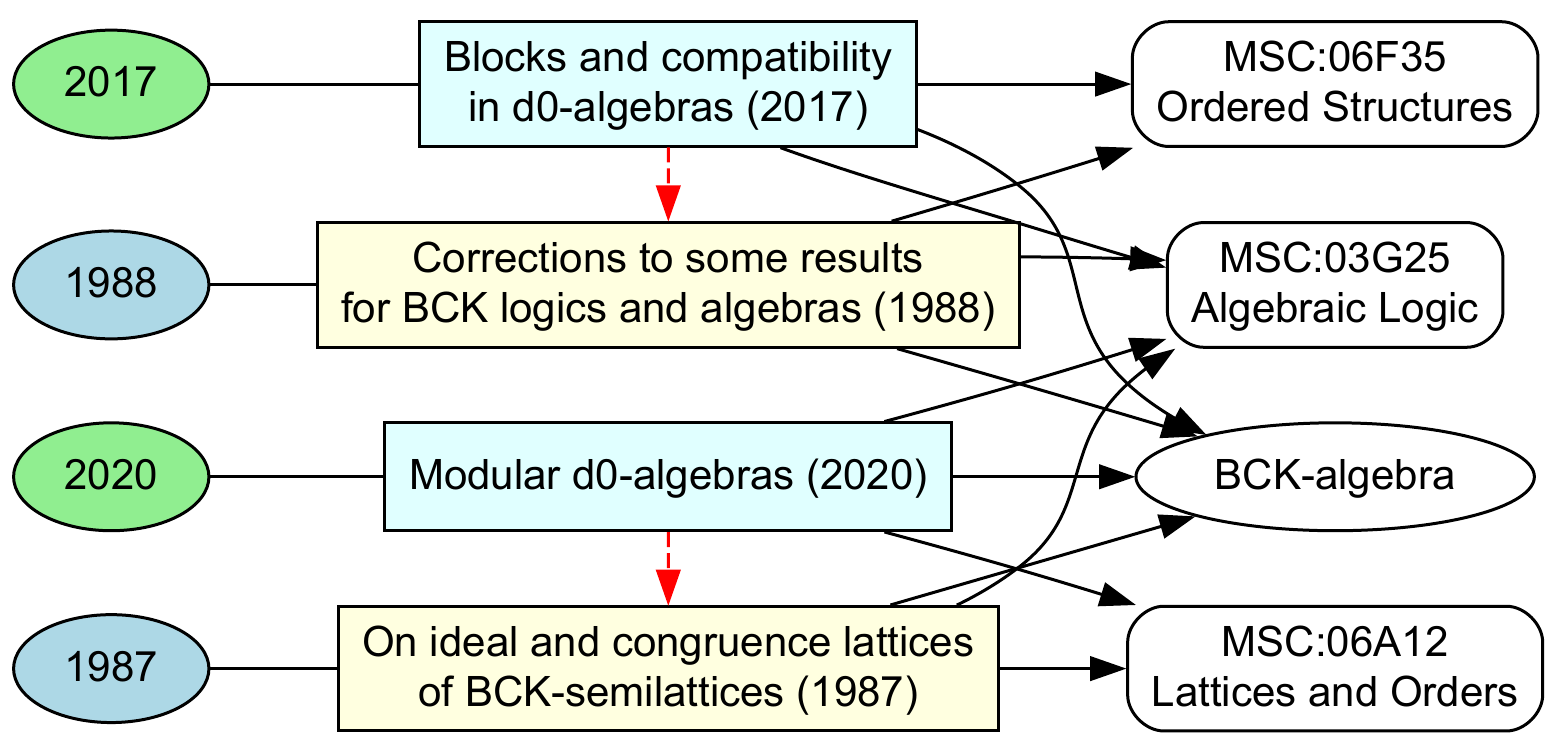}
    \caption{Potential connections identified through shared concepts and subject classifications.}
    \label{fig:prec}
\end{figure}
}

As an illustrative example of this use case, we queried \zb KG for the 2020 paper \textit{``Modular \( d_0 \)-algebras''} and identified the 1987 paper \textit{``On ideal and congruence lattices of BCK-semilattices''} as potential precursor connection. Although separated by 33 years, both works share subject classification codes (\texttt{MSC:03G25}, \texttt{MSC:06A12}) and the keyword \textit{BCK-algebra}. The earlier work investigates structural properties of BCK-semilattices, while the later work investigates \(d_0\)-algebras in a related mathematical setting.
Although the two publications are not explicitly connected by citation, their shared subject classifications and keywords, which were expert-curated, reveal a conceptual connection that may otherwise be difficult to identify through citation-based analysis. 
The knowledge graph surfaces the 1987 publication as a candidate for further scholarly and historical investigation. Other examples of such cases are illustrated in Figure~\ref{fig:prec}.

\subsection{Cross-Field Conceptual Continuity via MSC-Prefixes Bridging}
\label{sec:concept-ancestry}

The \zb KG enables exploration of how mathematical concepts recur across research fields through the hierarchical structure of the subject classification i.e., the MSC code. MSC codes provide a relatively stable taxonomy of mathematical subjects~\cite{msc2020} and, in the \zb KG, are assigned to the publication by mathematical experts. Such cross-field conceptual continuity can be difficult to identify from citation networks, as emerging fields may reinterpret established concepts within new frameworks without directly citing relevant earlier work. The \zb KG supports the systematic identification of recurring concepts across different mathematical fields and periods by leveraging MSC code hierarchies with controlled keywords and temporal information.

{
\setlength{\textfloatsep}{4pt} 
\setlength{\floatsep}{3pt}      
\setlength{\intextsep}{3pt}     
\begin{figure}
    \centering
    \includegraphics[width=1\linewidth]{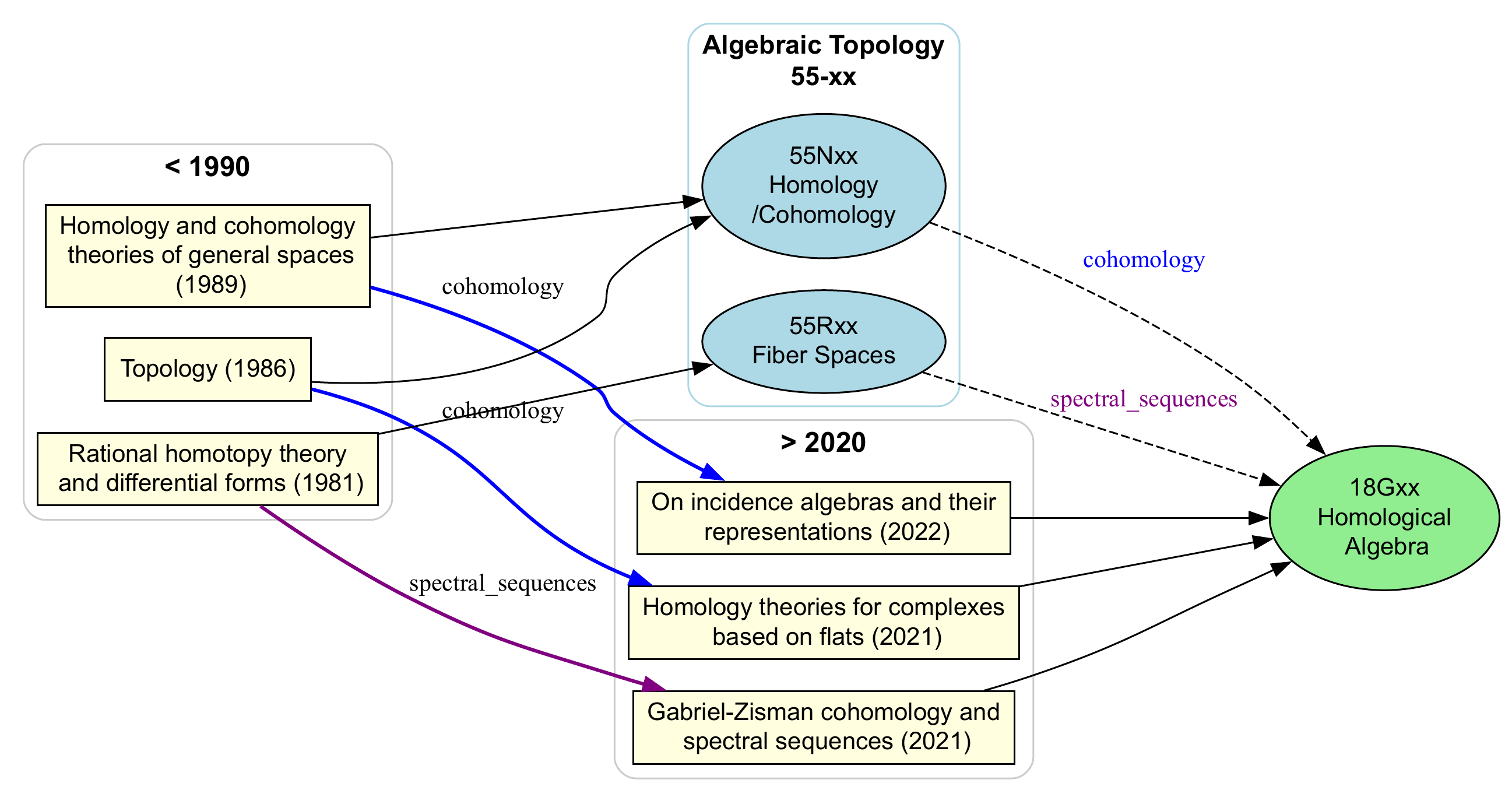}
    \caption{Conceptual Continuity: \textit{Algebraic Topology} $\to$ \textit{Homological Algebra}.}
    \label{fig:anc}
\end{figure}
}

An illustrative example derived from \zb KG examines the distribution of the concept \textit{spectral sequences} across two MSC areas: \textit{Algebraic Topology} (\code{MSC:55}) and \textit{Homological Algebra} (\code{MSC:18}), as illustrated in Figure~\ref{fig:anc}. Within Algebraic Topology, publications associated with the keyword occur in subfields such as homology and cohomology theories (\code{55Nxx}) and fiber spaces (\code{55Rxx}), where spectral sequences constitute an important methodological theme. The same keyword also occurs in later publications classified under Homological Algebra (\code{18Gxx}), illustrating its recurrence in a different disciplinary context. For example, the 2021 work \textit{``Gabriel-Zisman cohomology and spectral sequences''} is associated with the keyword and classified within Homological Algebra.

Using the keyword \textit{spectral sequences} together with temporal information and MSC classification codes, we identified a total of 463 early--later publication pairs spanning these two MSC areas. These pairs reveal recurring semantic associations across mathematical fields and time periods and provide candidates for further scholarly investigation of cross-field conceptual development. 

\subsection{Conceptual Resurgence: Tracking the Re-emergence of Ideas}
\label{sec:concept-revival}

Scientific concepts and ideas do not always develop along continuous trajectories. Some may receive sustained attention, decline in prominence, and later reappear in new theoretical or applied contexts. Identifying such patterns can provide historians with candidates for investigating intellectual cycles and help scientists discover areas experiencing renewed activity. The extensive temporal coverage of the \zb KG supports this exploration by combining temporal information with curated subject classifications and keywords, allowing patterns of emergence, decline, and resurgence to be examined systematically.

{
\setlength{\textfloatsep}{5pt} 
\setlength{\floatsep}{3pt}      
\setlength{\intextsep}{3pt}     
\begin{figure}
    \centering
    \includegraphics[width=0.9\linewidth]{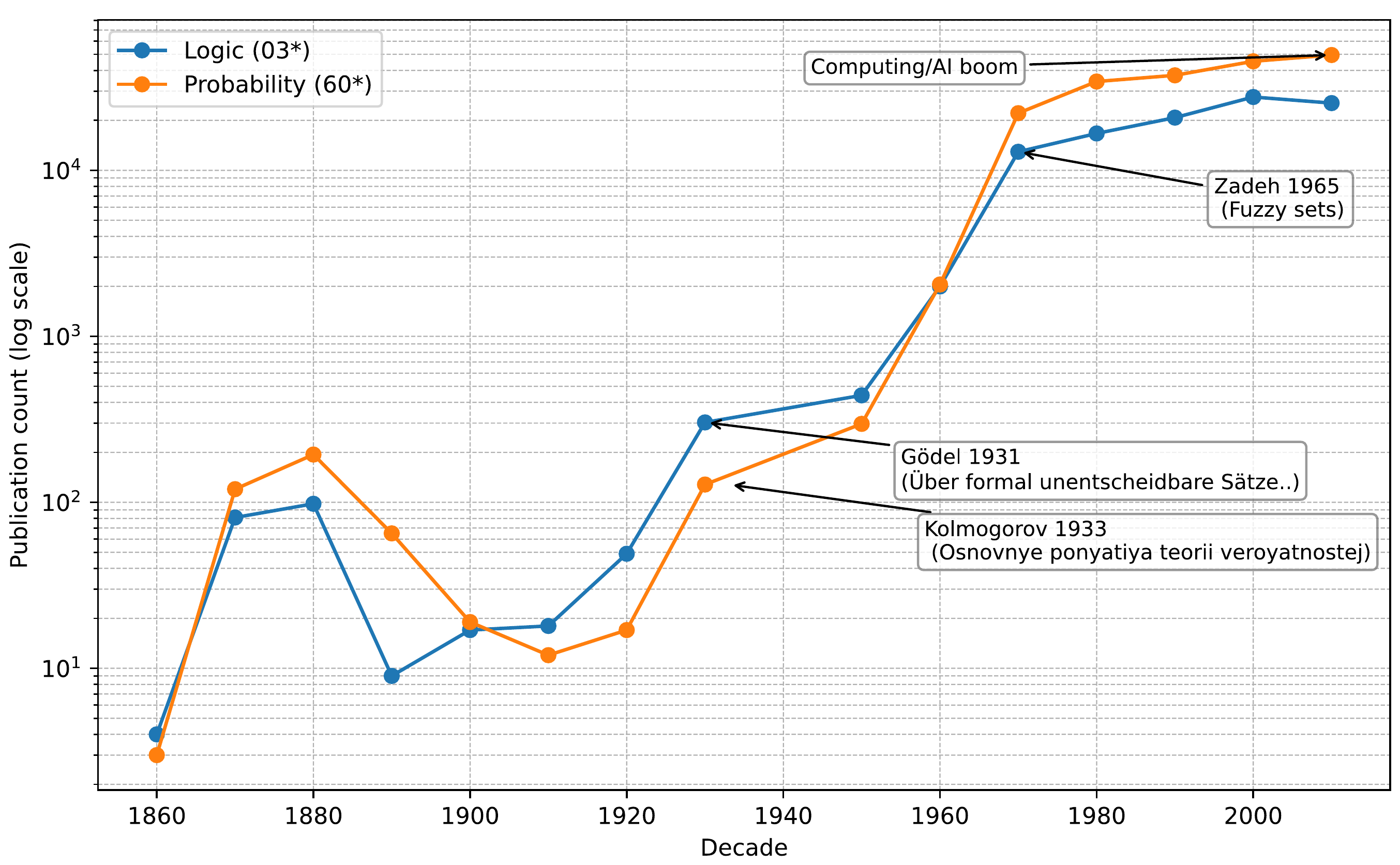}
    \caption{Temporal dynamics of mathematical subject areas: publication activity in \textit{Probability} (\code{MSC 60*}) and \textit{Logic} (\code{MSC 03*}) by decade.}
    \label{fig:revival}
\end{figure}
}

We queried \zb KG to illustrate this use case by examining publication activity in \textit{Probability} (\code{MSC:60}) and \textit{Logic} (\code{MSC:03}). Figure~\ref{fig:revival} presents the results aggregated by decade. 
Probability shows a sustained increase in publication activity throughout the twentieth century, with more pronounced growth after the 1950s. Representative works associated with this area include Kolmogorov's 1933 \textit{Grundbegriffe der Wahrscheinlichkeitsrechnung''}. 
Logic exhibits a marked increase during the 1930s (representative publication: Gödel’s incompleteness theorems), followed by a period of comparatively lower activity and renewed growth from the 1970s onward. Publications in this later period include works associated with fuzzy logic, such as Zadeh's 1965 \textit{Fuzzy Sets''}, as well as developments connecting logic with computer science and formal methods.
These patterns illustrate how the knowledge graph can make non-linear research dynamics visible at the subject-field level, identifying periods of changing research activity that can serve as starting points for further investigation of the concepts, methods, and publications associated with these transitions.

\subsection{Potential Intellectual Connections via Scholarly Engagement}
\label{sec:authorship}
The \zb KG incorporates mathematicians-curated reviews of publications, linking such reviewers to the publications they assessed and their authors through
\textit{reviewer--author} relationships.
These connections
provide an additional perspective for exploring potential intellectual continuities between a scholar’s earlier scholarly engagement as a reviewer and their subsequent research.

{
  \setlength{\textfloatsep}{3pt} %
  \setlength{\floatsep}{3pt}      %
  \setlength{\intextsep}{3pt}     %
\begin{figure}
    \centering
    \includegraphics[width=0.95\linewidth]{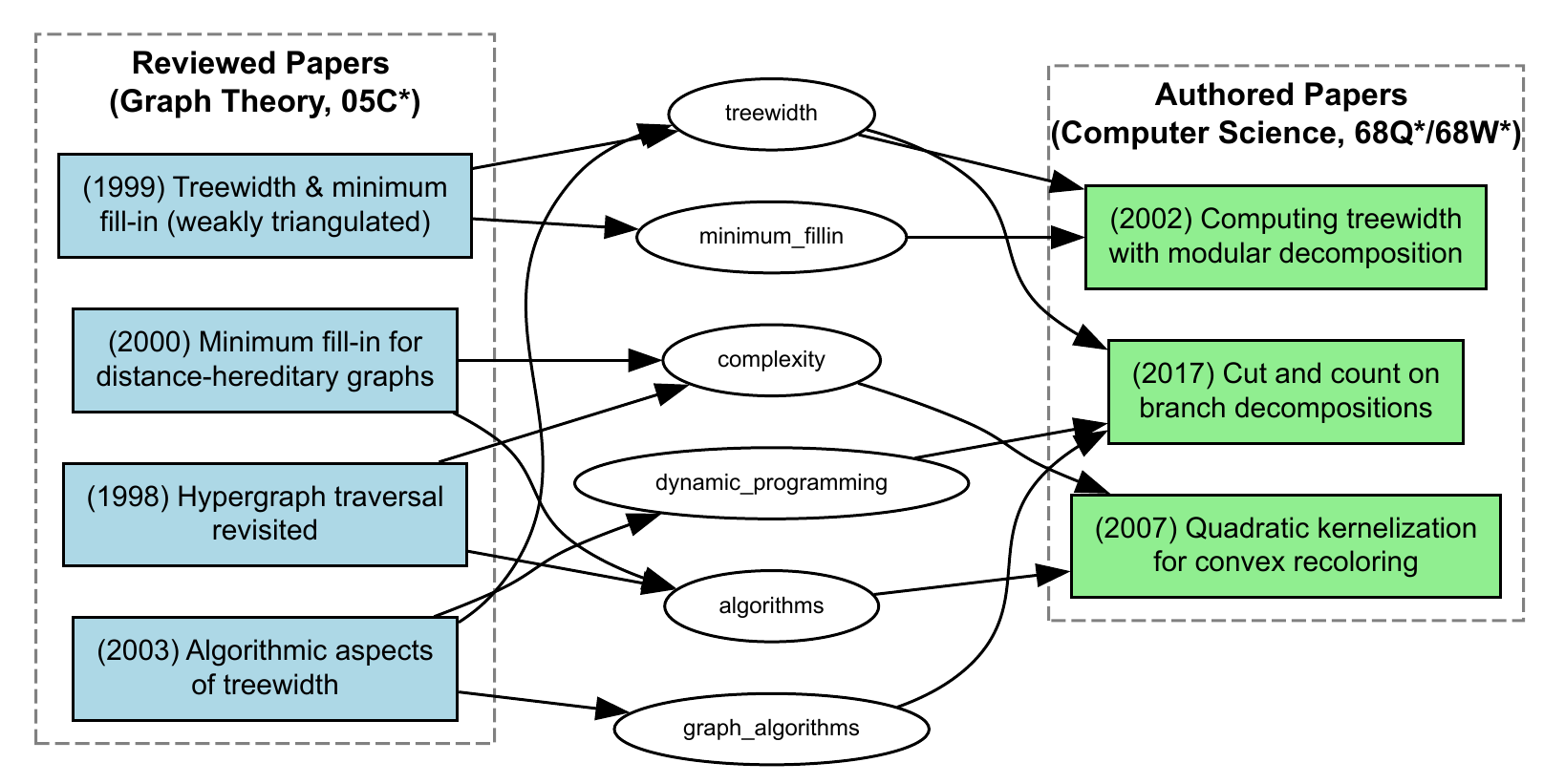}
    \caption{Potential intellectual continuity identified through author--reviewer relationships: the case of \textit{Hans L. Bodlaender}.}
    \label{fig:lineages}
\end{figure}
}

As an example, we queried the \zb KG for a specific scholar, \textit{Hans L. Bodlaender}.
We identified a potential intellectual connection between his early reviews (2000 or earlier) of graph-theoretic publications (\code{MSC 05C*}) and his later authored work (2000--2025) in the field of computer science (\code{MSC 68*}).
Keywords such as \textit{treewidth}, \textit{dynamic programming}, and \textit{graph algorithm} recur across the reviewed and authored publications, indicating topical overlap between research he encountered through reviewing in early years and topics appearing in his later authored publications. These overlaps identify candidate connections that can be further investigated using bibliographic and historical evidence. 
This example illustrates how \zb KG can complement publication- and citation-based analysis by incorporating human scholarly engagement into the exploration of potential research continuities and possible pathways of scholarly knowledge transmission.

\subsection{Other Potential Applications}
\label{sec:app}

The \textsf{zbMATH Open KG} provides an open semantic infrastructure for mathematical scholarship that can support applications beyond
bibliographic search and citation network-based analysis. Other potential applications include:

\paragraph{Historically Aware Scholarly Information Retrieval.}
The \zb KG supports scholarly retrieval that accounts for the historical development of mathematical knowledge over periods of time, supported by its extensive temporal coverage combined with expert-curated semantic information. It can potentially identify semantically related works despite changes in terminology, disciplinary boundaries, or the absence of direct citation links. This provides a basis for historically informed literature discovery that complements conventional keyword- and citation-based scholarly search.

\paragraph{Knowledge Development Analytics.}
The explicit representation of concepts through controlled keywords, subject domains through MSC codes, and expert-curated reviews, software references, disambiguated scholars, and their relationships, combined with extensive temporal coverage, provides a basis for quantitative analyses of mathematical knowledge over extended periods.
Researchers can investigate patterns of emergence, specialization, cross-field continuity, resurgence, and changing research activity, as well as long-term trends in the development of mathematical disciplines over more than 250 years. 

\paragraph{Semantic Infrastructure for AI System.}
The \zb KG can serve as a semantic backbone for AI-assisted scholarly applications, such as graph-based scientific question answering and recommendation systems~\cite{dasfa25,doweneed26}, retrieval-augmented generation, and agentic literature exploration. Particularly, explicit semantic relationships via the hierarchical MSC codes and controlled keywords provide structured sources of context that complement bibliographic metadata and unstructured text, while expert-curated reviews add broader contextual interpretations of publications beyond its original text. This supports more grounded and explainable AI-assisted exploration of mathematical literature.

\vspace{-2mm}
\section{FAIR Compliance and Sustainability}
\label{sec:fair}
\paragraph{Findable} Persistent URIs are assigned to all entities, with URI resolution enabled for the core entities (e.g., publications, scholars, software references). RDF dump releases are versioned and accompanied by machine-readable metadata describing the scope, schema, and statistics of the corresponding release. Dataset-level descriptions (VoID, DCAT, PROV-O) support machine-readable discovery and indexing within the Linked Data ecosystem.

\paragraph{Accessible} The complete RDF data are distributed as downloadable dumps through Zenodo, providing a persistent distribution channel and long-term preservation of released snapshots. The RDF representation can be loaded into standard RDF triple stores; the pipeline has been tested with Virtuoso and Apache Jena. A public SPARQL endpoint is provided to support direct querying of the published knowledge graph in addition to bulk access through RDF dumps.

\paragraph{Interoperable} 
The KG uses W3C Semantic Web standards and systematically reuses established vocabularies.
Where available, external identifiers such as DOIs and other external links and identifiers are retained, facilitating linking with external scholarly resources and knowledge graphs. 

\paragraph{Reusable} The resource is accompanied by its OWL ontology and data-level description, RDF dumps, documentation, and source code. Stable namespaces and identifiers support the reuse of entities across releases; versioned RDF snapshots enable reproducible experiments against a fixed resource state. Licensing and attribution information for the released data and accompanying artifacts is provided with each distribution. The released knowledge graph and all accompanying artifacts and source code are distributed under CC-BY-SA 4.0.

\paragraph{Sustainability and Maintenance} 
The knowledge graph construction process is designed as a reproducible pipeline over the continuously maintained zbMATH Open platform. Periodic regeneration from updated source data, versioned releases, persistent entity identifiers, and public repository hosting support the continued evolution of the knowledge graph while allowing individual releases to remain independently reproducible. The ontology is maintained separately from individual data snapshots, enabling future extensions without requiring changes to the underlying identifier scheme. RDF dump releases are updated periodically, approximately twice per year subject to the availability of the underlying data and resource and infrastructure constraint.

\section{Conclusion}
\label{sec:conclusion}
We presented \zb Knowledge Graph, a large scale RDF knowledge graph that integrates more than 250 years of mathematical scholarship grounded in expert-curated semantic information. Its extensive temporal coverage and domain-specific semantic structure provide a reusable foundation for computational exploration of mathematical knowledge beyond conventional bibliographic and citation-based analysis. The \zb KG is openly available and designed to align with FAIR principles and Semantic Web standards.

We demonstrated the utility of the knowledge graph through historically grounded scholarly exploration use cases, illustrating how it can identify patterns and candidate relationships that may be difficult to uncover in conventional citation networks and provide starting points for further scholarly and historical investigation.
Future directions including, but not limited to:
\begin{itemize}[itemsep=5pt]
    \item[(i)] 
    Extending the \zb KG through links to external knowledge graphs and related resources e.g., linkage to \textsf{SemRepo}\cite{semrepo} could enrich the current mathematical software references, enabling the study of the development and adoption of mathematical software and infrastructure over time. 

    \item[(ii)] 
    Investigating how KG-based context from \zb KG complements textual and embedding-based methods in AI-assisted retrieval and exploration of the mathematical literature.
\end{itemize}

\begin{credits}
\subsubsection{Supplementary Material:} See Page~\pageref{firstpage}.

\subsubsection{Acknowledgment}
This work was supported by the LUMEN 
project, funded by the European Union's Horizon Europe Research and Innovation Programme (Grant: \href{https://cordis.europa.eu/project/id/101187940}{101187940}, and by the DFG`s MaRDI project (Grant: \href{https://gepris.dfg.de/project/460135501/2}{460135501}). The views and opinions expressed are those of the authors only and do not necessarily reflect those of the funding organizations or the granting authorities. 

\subsubsection{\discintname}
The authors have no competing interests to declare.

\subsubsection*{Declaration of use of Generative AI.}
We utilized generative AI to support the writing process throughout this work. All AI-generated content was carefully reviewed, edited, and approved by the authors, who assume full responsibility of the final manuscript.

\end{credits}
\bibliographystyle{splncs04}
\bibliography{ref}
\end{document}